\documentclass{article}
\usepackage{spconf,amsmath,graphicx,booktabs,multirow}

\usepackage{enumitem}
\setlist{nosep, leftmargin=14pt}

\newcommand{\customsmall}{\fontsize{8.5}{10}\selectfont}

\title{Bayesian Topological Analysis of Functional Brain Networks}

\name{Xukun Zhu$^{1}$, Michael W Lutz$^2$, Tananun Songdechakraiwut$^{3}$}

\address{
$^{1}$Department of Statistical Science, Duke University \\
$^{2}$Departments of Neurology and Pathology, Duke University School of Medicine \\
$^{3}$Department of Computer Science, Duke University 
}

\begin{document}

\maketitle

\begin{abstract}
Subtle alterations in brain network topology often evade detection by traditional statistical methods. To address this limitation, we introduce a Bayesian inference framework for topological comparison of brain networks that probabilistically models within- and between-group dissimilarities. The framework employs Markov chain Monte Carlo sampling to estimate posterior distributions of test statistics and Bayes factors, enabling graded evidence assessment beyond binary significance testing. Simulations confirmed statistical consistency to permutation testing. Applied to fMRI data from the Duke-UNC Alzheimer's Disease Research Center, the framework detected topology-based network differences that conventional permutation tests failed to reveal, highlighting its enhanced sensitivity to early or subtle brain network alterations in clinical neuroimaging.

\end{abstract}


\section{Introduction}

Persistent homology~\cite{edelsbrunner2022computational} has emerged as a powerful framework for characterizing brain network topology in neurodegenerative diseases, capturing multi-scale network organization through persistence diagrams and Wasserstein distance metrics~\cite{yi2025topological}. To statistically compare these topological representations across clinical groups, permutation testing remains the standard approach for group comparisons in neuroimaging, providing non-parametric inference without distributional assumptions~\cite{nichols2002nonparametric}. However, permutation tests \cite{songdechakraiwut2023topological} yield only binary significance decisions at predetermined thresholds and require computationally intensive resampling for large datasets.

Bayesian Statistical inference~\cite{hoff2009first} offers a complementary framework that addresses these limitations. By probabilistic modeling of network dissimilarities, Bayesian methods provide complete posterior distributions rather than point estimates, enabling evidence assessment through Bayes factors that quantify the strength of evidence for competing hypotheses and uncertainty quantification via credible intervals~\cite{heckerman1998tutorial,kass1995bayes}. However, despite these methodological advantages, Bayesian approaches remain underutilized for topological group comparisons in functional brain networks.

The main contributions of this paper are as follows. 1) We present a Bayesian inference framework for comparing brain network topologies that models topological dissimilarities and provides posterior distributions, Bayes factors, and credible intervals. 2) We validate the proposed framework using simulation experiments, confirming its statistical consistency with permutation testing. 3) The method is applied to resting-state fMRI data from the Duke-UNC Alzheimer's Disease Research Center, demonstrating enhanced sensitivity in detecting amyloid and tau pathology-related network alterations compared to conventional permutation testing.

\section{Methods}

\subsection{Topological Distance and Group Dissimilarity}

Each brain network is represented as an undirected, weighted graph derived from functional connectivity. We employ persistent graph homology (PGH), which captures interpretable topological invariants, connected components (0-homology) and independent cycles (1-homology), across the entire range of edge weights using closed-form computation \cite{songdechakraiwut2023wasserstein}. Given a functional connectivity graph $G = (V, W)$, where $V$ denotes the set of brain regions and $W = \{w_{ij}\}$ represents pairwise functional connections, we define a sequence of binary graphs $G_{\epsilon}$ by applying thresholds to the weights: an edge $(i,j)$ exists in $G_{\epsilon}$ if $w_{ij} > \epsilon$. As $\epsilon$ increases, a nested sequence (filtration) of graphs is formed:
$G_{\epsilon_0} \supseteq G_{\epsilon_1} \supseteq \cdots \supseteq G_{\epsilon_k},$
where $\epsilon_0 \leq \epsilon_1 \leq \cdots \leq \epsilon_k$ are filtration values. PGH tracks how topological features emerge and disappear across filtration levels. A feature born at $b_l$ and dying at $d_l$ is represented by a point $(b_l, d_l)$ in a persistence diagram \cite{edelsbrunner2022computational}. As $\epsilon$ increases, the number of connected components $\beta_0$ increases monotonically, while the number of cycles $\beta_1$ decreases monotonically. This monotonic behavior ensures that it suffices to record only the birth times of connected components and the death times of cycles:
$I_0(G) = \{b_l\}_{l=1}^{|V|-1}, 
I_1(G) = \{d_l\}_{l=1}^{1 + |V|(|V| - 3)/2}.$ Therefore, we represent the persistence diagram using birth sets for connected components ($I_0$) and death sets for cycles ($I_1$) \cite{songdechakraiwut2021topological}.

The topological distance between two networks $G_1$ and $G_2$ is defined as the combined Wasserstein distance across graph homology dimensions:
\begin{equation}
    \widetilde{W}_{\text{top}}(G_1, G_2) = W_{0D}(G_1, G_2) + W_{1D}(G_1, G_2),
\end{equation}
where $W_{0D}$ and $W_{1D}$ measure dissimilarity in terms of connected components and cycles, respectively. 
The Wasserstein distance for connected components is defined as:
\begin{equation}
    W_{0D}(G_1, G_2) = \min_{\tau_0} \sum_{b \in I_0(G_1)} \|b - \tau_0(b)\|^2,
\end{equation}
where $\tau_0$ is a bijection between the birth sets $I_0(G_1)$ and $I_0(G_2)$. Similarly, the Wasserstein distance for cycles is defined as:
\begin{equation}
    W_{1D}(G_1, G_2) = \min_{\tau_1} \sum_{d \in I_1(G_1)} \|d - \tau_1(d)\|^2,
\end{equation}
where $\tau_1$ is a bijection between the death sets $I_1(G_1)$ and $I_1(G_2)$.

Let $\mathcal{X} = \{X_1, \dots, X_m\}$ and $\mathcal{Y} = \{Y_1, \dots, Y_n\}$ denote two groups of brain networks (e.g., Alzheimer's disease patients and healthy controls). We quantify group-level dissimilarity by computing pairwise topological distances between networks. The average within-group distance is defined as:
\begin{equation}
    \bar{D}_W = \frac{\sum_{i<j} \widetilde{W}_{\text{top}}(X_i, X_j) + \sum_{i<j} \widetilde{W}_{\text{top}}(Y_i, Y_j)}{\binom{m}{2} + \binom{n}{2}},
\end{equation}
which measures the typical topological variability within homogeneous groups. The average between-group distance is:
\begin{equation}
    \bar{D}_B = \frac{1}{mn} \sum_{i=1}^{m} \sum_{j=1}^{n} \widetilde{W}_{\text{top}}(X_i, Y_j),
\end{equation}
reflecting the topological distinction between clinical populations. We use the individual pairwise distances $d^{\text{within}}_i = \widetilde{W}_{\text{top}}(X_a, X_b)$ for all pairs within groups and $d^{\text{between}}_j = \widetilde{W}_{\text{top}}(X_i, Y_j)$ for all cross-group pairs as observed data, rather than the averaged quantities $\bar{D}_W$ and $\bar{D}_B$.

\subsection{Bayesian Inference Framework}

Given the observed topological distances $\mathbf{d} = \{d^{\text{within}}_i, d^{\text{between}}_j\}$ from Section 2.1, we propose a Bayesian inference framework to quantify uncertainty in group differences. Unlike permutation tests that compare point estimates $\bar{D}_B$ and $\bar{D}_W$ using empirical null distributions, our approach yields complete posterior distributions of test statistics, enabling probabilistic statements about the magnitude and reliability of group differences. To ensure numerical stability and satisfy distributional assumptions, we standardize distances separately within each group prior to modeling, then back-transform posterior estimates to the original Wasserstein distance scale, ensuring all reported statistics retain direct interpretability in terms of topological dissimilarity.We model the within-group and between-group distances as normal distributions:
\begin{align}
d^{\text{within}}_i &\sim \mathcal{N}(\mu_{\text{w}}, \sigma_{\text{w}}^2), \quad i = 1, \ldots, n_{\text{w}} \\
d^{\text{between}}_j &\sim \mathcal{N}(\mu_{\text{b}}, \sigma_{\text{b}}^2), \quad j = 1, \ldots, n_{\text{b}}
\end{align}
where $\mu_{\text{w}}, \mu_{\text{b}}$ represent mean distances and $\sigma_{\text{w}}^2, \sigma_{\text{b}}^2$ are variance. The parameter vector $\boldsymbol{\theta} = (\mu_{\text{w}}, \mu_{\text{b}}, \sigma_{\text{w}}^2, \sigma_{\text{b}}^2)$ is treated as random with prior distributions $p(\boldsymbol{\theta})$. We adopt weakly informative priors that regularize inference without imposing strong assumptions:
\begin{equation}
\mu_{\text{w}}, \mu_{\text{b}} \sim \mathcal{N}(0, 10), \quad
\sigma_{\text{w}}^2, \sigma_{\text{b}}^2 \sim \text{InvGamma}(0.01, 0.01)
\end{equation}

By Bayes' theorem, the posterior distribution is $p(\boldsymbol{\theta} \mid \mathbf{d}) \propto p(\mathbf{d} \mid \boldsymbol{\theta}) p(\boldsymbol{\theta})$, where the likelihood factorizes as:
\begin{equation}
p(\mathbf{d} \mid \boldsymbol{\theta}) = \prod_{i=1}^{n_{\text{w}}} \mathcal{N}(d^{\text{within}}_i \mid \mu_{\text{w}}, \sigma_{\text{w}}^2) \prod_{j=1}^{n_{\text{b}}} \mathcal{N}(d^{\text{between}}_j \mid \mu_{\text{b}}, \sigma_{\text{b}}^2)
\end{equation}
We perform Markov Chain Monte Carlo (MCMC)  sampling to obtain $S$ posterior samples $\{\boldsymbol{\theta}^{(s)}\}_{s=1}^{S}$. From these samples, we derive distributions of two test statistics: the ratio $\phi_L = \mu_{\text{b}} / \mu_{\text{w}}$ measuring relative dissimilarity , which is also used in permutation testing. And the mean difference $\delta = \mu_{\text{b}} - \mu_{\text{w}}$ quantifying group separation. For each posterior sample $s$, we compute $\delta^{(s)} = \mu_{\text{b}}^{(s)} - \mu_{\text{w}}^{(s)}$ and $\phi_L^{(s)} = \mu_{\text{b}}^{(s)} / \mu_{\text{w}}^{(s)}$. The posterior distributions provide complete uncertainty quantification, from which we extract posterior means and 95\% credible intervals for both parameters and test statistics.

Furthermore, to quantify the strength of evidence for group differences, we compute the Bayes Factor comparing the alternative hypothesis $H_1$ (group differences exist, $\mu_{\text{b}} \neq \mu_{\text{w}}$) against the null hypothesis $H_0$ (no group differences, $\mu_{\text{b}} = \mu_{\text{w}}$):
\begin{equation}
BF_{10} = \frac{p(\mathbf{d} \mid H_1)}{p(\mathbf{d} \mid H_0)}
\end{equation}
The Bayes Factor is computed via Leave-one-out cross-validation (LOO-CV) as the log difference between expected log-predictive densities: $\log BF_{10} = \text{elpd}_{H_1} - \text{elpd}_{H_0}$.
Following Jeffreys' classification \cite{jeffreys1998theory}, we interpret the strength of evidence based on $BF_{10}$ values: $BF_{10} < 1$ supports $H_0$, with $BF_{10} < 0.1$ indicating strong evidence; $BF_{10} > 1$ supports $H_1$, with $BF_{10} > 10$ considered strong and $BF_{10} > 100$ decisive evidence. Unlike traditional hypothesis testing, we accept or reject the null hypothesis by comparing p-value with fixed thresholds, the Bayes Factor quantifies not only the results but also strength of evidence for hypothesis testing. This provides more detailed interpretations, particularly when p-values are close to significance thresholds where simple accept/reject conclusions may be insufficient.

\begin{table}[t]
\centering
\customsmall 
\setlength{\tabcolsep}{0.5pt}  
\begin{tabular}{l l c c}
\toprule
\textbf{Method} & \textbf{Statistic} & \textbf{Exp 1: Different} & \textbf{Exp 2: Same} \\
\midrule
\multirow{3}{*}{\textit{Permutation Test}} 
& Observed $\phi$ & 7.022 & 0.946 \\
& $p$-value & $<0.001$ & 0.634 \\
& Decision & Significant & Not Sig. \\
\midrule
\multirow{3}{*}{\textit{Bayesian Ratio ($\phi$)}} 
& Posterior Mean & 7.120 & 0.953 \\
& 95\% CI & [5.592, 9.261] & [0.734, 1.223] \\
& $P(\phi>1)$ & 1.000 & 0.330 \\
\midrule
\multirow{3}{*}{\textit{Bayesian Diff. ($\Delta$)}} 
& Posterior Mean & 16.156 & $-0.192$ \\
& 95\% CI & [14.724, 17.585] & [$-1.042$, 0.669] \\
& $P(\Delta>0)$ & 1.000 & 0.330 \\
\midrule
\multirow{2}{*}{\textit{Bayes Factor}} 
& $\text{BF}_{10}$(LOO) & $1.95 \times 10^{139}$ & $6.46 \times 10^{-3}$ \\
& Interpretation & Decisive for $H_1$ & Moderate for $H_0$ \\
\bottomrule
\end{tabular}
\caption{Validation results comparing permutation testing and Bayesian inference. Experiment 1 tested networks with different structures ($c = 2$ vs $c = 3$), while Experiment 2 tested networks with identical structures ($c = 3$ vs $c = 3$).}
\label{tab:case1_summary}
\end{table}

\subsection{Simulation-Based Validation}

To validate the equivalence between the Bayesian framework and permutation testing, we conducted simulation experiments under two cases using modular networks with $n=24$ nodes generated via stochastic block models: networks with different structures (c=2 vs c=3) and networks with identical structures (c=3 vs c=3), with 10 networks per group \cite{songdechakraiwut2023topological}. While real brain networks may exhibit varying degrees of similarity, these conditions allow us to assess the framework's ability to correctly identify both presence and absence of topological differences \cite{bullmore2009complex}. Connection probabilities were $p_{\text{within}}=0.7$ and $p_{\text{between}}=0.1$, and permutation testing used 5,000 transpositions while Bayesian models employed MCMC with 4 chains and 8,000 draws. As shown in Table~\ref{tab:case1_summary}, both methods are statistically consisitent: for different structures, permutation test yielded $p<0.001$ and Bayesian analysis gave $P(\phi>1)=1.000$ with decisive evidence; for identical structures, both correctly identified no significant difference ($p=0.634$, $P(\phi>1)=0.330$), with Bayes factor uniquely quantifying moderate evidence for the null hypothesis.

\begin{table}[t!]
\centering
\customsmall  
\setlength{\tabcolsep}{2.5pt} 
\begin{tabular}{l l c c}
\toprule
\textbf{Method} & \textbf{Statistic} & \textbf{A+ vs A-} & \textbf{T+ vs T-} \\
\midrule
\multirow{3}{*}{\textit{Permutation Test}} 
& Observed $\phi$ & 1.103 & 1.369 \\
& $p$-value & 0.221 & 0.036 \\
& Decision & Not Sig. & Significant \\
\midrule
\multirow{3}{*}{\textit{Bayesian Ratio ($\phi$)}} 
& Posterior Mean & 1.104 & 1.369 \\
& 95\% CI & [1.048, 1.161] & [1.296, 1.443] \\
& $P(\phi>1)$ & 1.000 & 1.000 \\
\midrule
\multirow{3}{*}{\textit{Bayesian Diff. ($\Delta$)}} 
& Posterior Mean & 5.968 & 20.416 \\
& 95\% CI & [2.794, 9.118] & [16.655, 24.131] \\
& $P(\Delta>0)$ & 1.000 & 1.000 \\
\midrule
\multirow{2}{*}{\textit{Bayes Factor}} 
& $\text{BF}_{10}$ (LOO) & $1.33 \times 10^{5}$ & $3.89 \times 10^{27}$ \\
& Interpretation & Decisive for $H_1$ & Decisive for $H_1$ \\
\bottomrule
\end{tabular}
\caption{Statistical comparison across biomarker groups. Permutation testing misses amyloid differences ($p=0.221$) and marginally detects tau differences ($p=0.036$), while Bayesian analysis provides decisive evidence in both cases.}
\label{tab:results_combined}
\end{table}

\section{Application to Alzheimer's Disease}

\begin{figure}[t]
\begin{minipage}[b]{1.0\linewidth}
  \centering
  \centerline{\includegraphics[width=8.5cm, height=6.5cm]{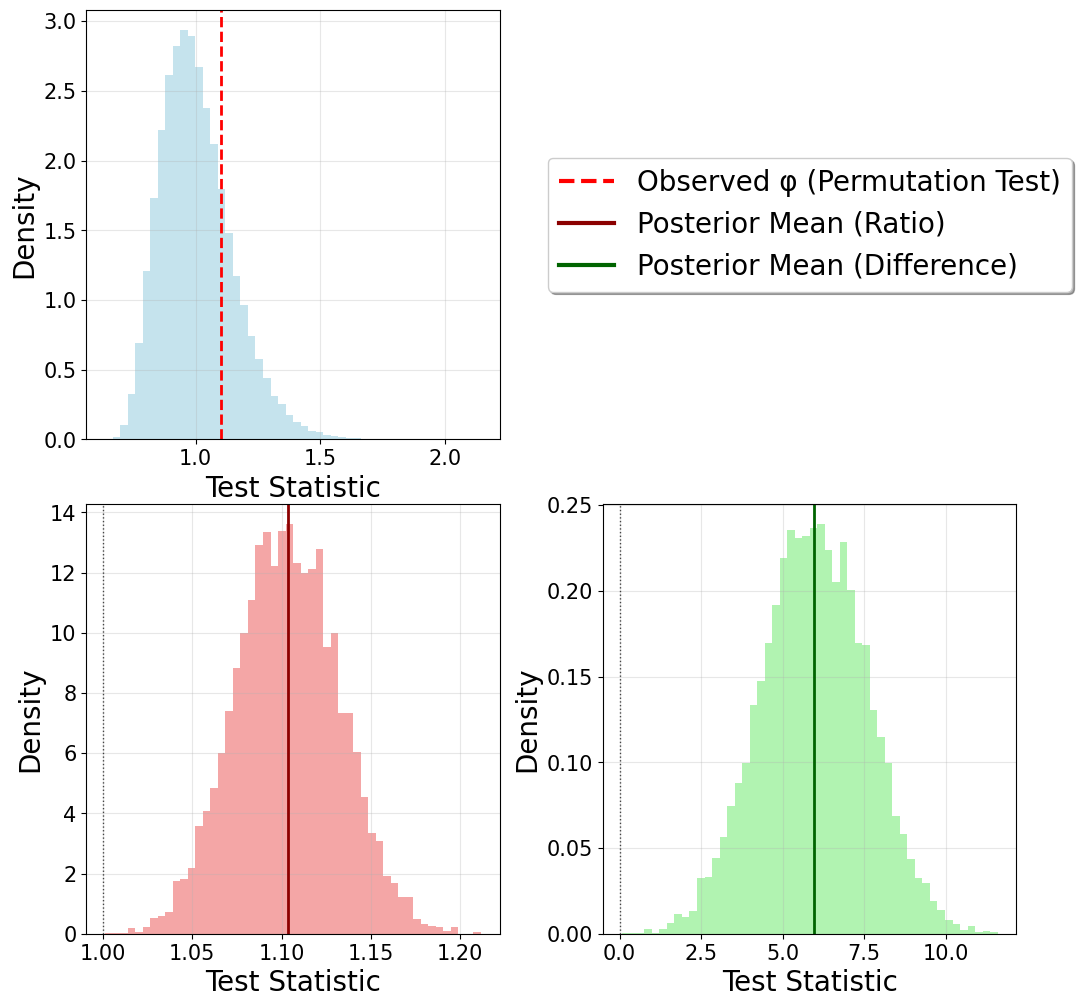}}
\end{minipage}
\caption{A+ vs A- comparison. Top: Permutation test null distribution (blue) with observed $\phi$ (red dashed line). Bottom left: Posterior distribution for ratio$\phi$ (red) with posterior mean (dark red line). Bottom right: Posterior distribution for difference $\Delta$ (green) with posterior mean (dark green line). Despite non-significant permutation results, Bayesian methods provides decisive evidence for topological differences between groups.}
\label{fig:amyloid_comparison}
\end{figure}

\begin{figure}[t]
\begin{minipage}[b]{1.0\linewidth}
  \centering
  \centerline{\includegraphics[width=8.5cm, height=6.5cm]{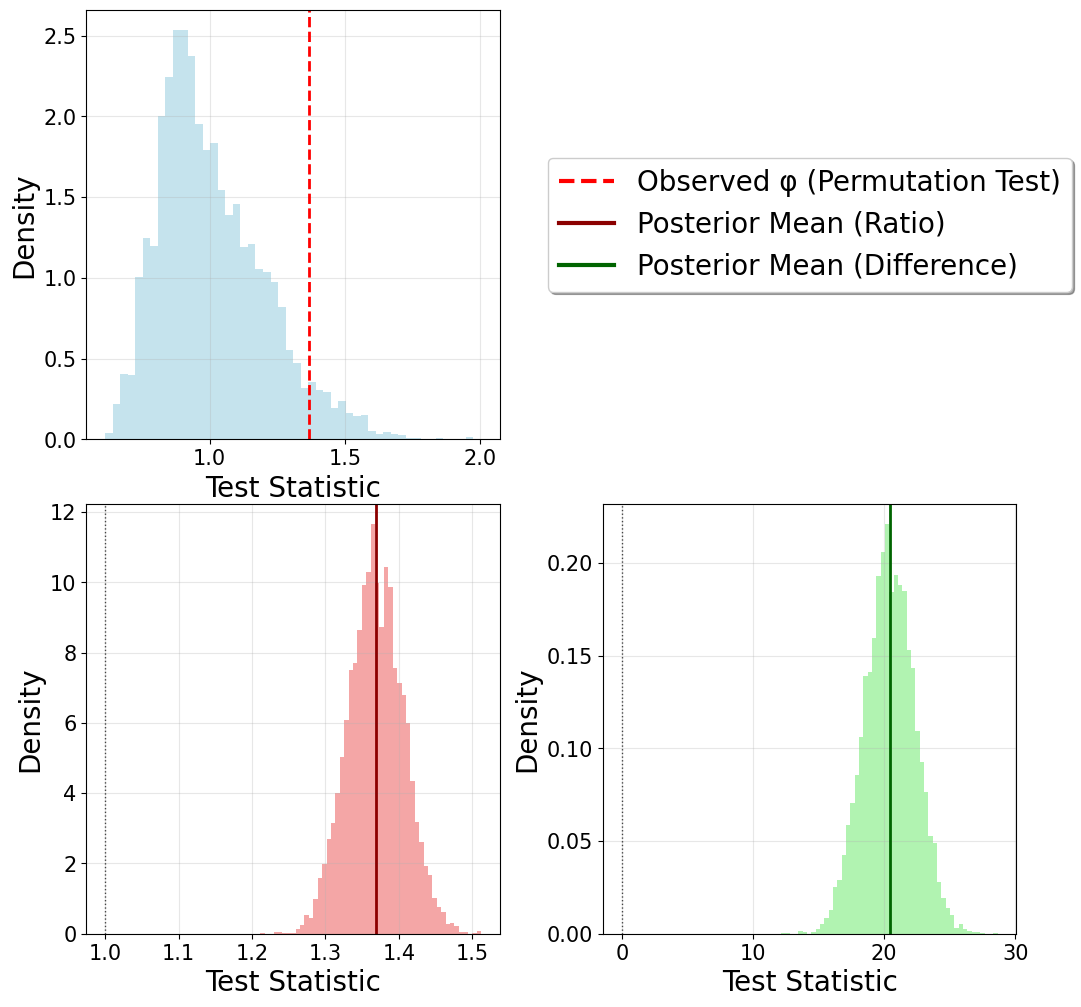}}
\end{minipage}
\caption{T+ vs T- comparison: Top: Permutation test null distribution (blue) with observed $\phi$ (red dashed line). Bottom left: Posterior distribution for ratio$\phi$ (red) with posterior mean (dark red line). Bottom right: Posterior distribution for difference $\Delta$ (green) with posterior mean (dark green line). Both permutation test and Bayesian methods demonstrate significant topological differences between groups.}
\label{fig:tau_comparison}
\end{figure}

Amyloid-beta and tau represent the core pathological hallmarks of Alzheimer's disease~\cite{knopman2021alzheimer}. Understanding how these biomarkers differentially affect brain network topology is critical for early detection and intervention strategies. We applied our framework to resting-state fMRI data from the Duke/UNC Alzheimer's Disease Research Center (ADRC). The Duke/UNC ADRC emphasizes age-related factors across the lifespan that contribute to AD and dementia, and many participants are relatively young and cognitively unimpaired. The dataset comprised 167 participants who underwent both neuroimaging and cerebrospinal fluid biomarker assessments. fMRI data were preprocessed using fMRIPrep with standard preprocessing steps, and brain networks were constructed using the AAL atlas (116 regions). Participants were classified by amyloid-beta (A) and tau (T) pathology using the AT(N) framework, yielding two comparisons: amyloid-positive (A+) versus amyloid-negative (A-), and tau-positive (T+) versus tau-negative (T-). The connectivity matrices were age-adjusted~\cite{damoiseaux2017effects} and analyzed using both permutation testing and Bayesian framework. Results are summarized in Table~\ref{tab:results_combined}. Figures~\ref{fig:amyloid_comparison} and~\ref{fig:tau_comparison} visualize the difference between these approaches: permutation testing evaluates the observed statistic against its null distribution under the no-difference hypothesis, while Bayesian methods directly characterizes the posterior distribution of statistics given the observed data.

\paragraph*{Amyloid status comparison.}
Permutation testing yielded $p=0.221$, failing to detect significant differences at the conventional $\alpha=0.05$ threshold. In contrast, Bayesian analysis shows $\phi=1.104$ (95\% CI: [1.048, 1.161]) with $P(\phi>1)=1.000$ and $\text{BF}_{10}=1.33 \times 10^{5}$, providing decisive evidence for group differences. This demonstrates the framework's enhanced sensitivity to subtle network alterations: the method detects approximately 10\% increase in between-group dissimilarity ($\phi=1.104$) and 6-unit absolute difference ($\Delta=5.968$) that threshold-based testing completely misses. Such sensitivity is particularly valuable for early-stage pathology like amyloid accumulation, where network changes are modest but may carry important clinical implications for disease progression.

\paragraph*{Tau status comparison}
Permutation testing yielded $p=0.036$, marginally achieving significance but just barely crossing the threshold. Bayesian analysis provides substantially stronger evidence with $\phi=1.369$ (95\% CI: [1.296, 1.443]), $P(\phi>1)=1.000$, and $\text{BF}_{10}=3.89 \times 10^{27}$ . Notably, the tau effect is substantially larger than amyloid: $\phi=1.369$ versus $1.104$ represents 37\% versus 10\% increases in between-group dissimilarity, while $\Delta=20.416$ versus $5.968$ shows a difference in distance. This quantitative pattern is consistent with neuroscience literature indicating tau pathology, as a later-stage process, produces more severe functional network disruption than early amyloid accumulation.

\section{Discussion}

We demonstrated that resting-state fMRI exhibits detectable topological alterations associated with amyloid-beta and tau pathology in Alzheimer's disease. Tau pathology produces substantially larger network disruptions than amyloid, consistent with literature indicating tau represents later-stage pathological process. Notably, the Bayesian approach detected amyloid-related differences permutation testing missed, suggesting enhanced sensitivity for early-stage changes.

The framework provides uncertainty quantification through posterior distributions and Bayes factors, enabling graded evidence assessment rather than binary decisions. This proves valuable when interpreting results near decision thresholds or with limited samples.

Several extensions of this framework merit future investigation. For instance, applying the method to region-specific connectivity could localize topological alterations to particular functional networks\cite{greicius2004default}, while incorporating longitudinal data may identify critical transition points in network degradation over disease progression.

\section{Compliance with Ethical Standards}

The ADRC study is approved by the Duke University and University of North Carolina at Chapel Hill institutional review boards in accordance with ethical principles and applicable regulatory requirements (Duke University IRB Protocol Number: Pro00103958). Written informed consent is obtained from all participants ages 25 to 80 years.

\section{Acknowledgements}

Research reported in this publication was supported by Duke Science and Technology LAUNCH grant, Bass Connections at Duke University, and the National Institute on Aging of the National Institutes of Health under Award Number P30AG072958.

\bibliographystyle{IEEEbib}
\bibliography{refs}

\end{document}